\documentclass[12pt, epsfig]{article}
\usepackage{amsmath,amsfonts,amssymb,amsthm,amstext,amscd,eucal}
\usepackage[all]{xy}
\usepackage{hyperref}
\usepackage[dvips]{color}
\usepackage{epsfig}%
\usepackage[active]{srcltx}
\makeatletter \@addtoreset{equation}{section}

\makeatletter\renewcommand\section{\@startsection {section}{1}{\z@}%
                                   {-3.5ex \@plus -1ex \@minus -.2ex}
                                   {2.3ex \@plus.2ex}%
                                   {\normalfont\large\bfseries}}
\renewcommand\subsection{\@startsection{subsection}{2}{\z@}%
                                     {-3.25ex\@plus -1ex \@minus -.2ex}%
                                     {1.5ex \@plus .2ex}%
                                     {\normalfont\bfseries}}
\newcommand{\be}{\begin{equation}}
\newcommand{\ee}{\end{equation}}
\newcommand{\bea}{\begin{eqnarray}}
\newcommand{\eea}{\end{eqnarray}}
\newcommand{\bse}{\begin{subequations}}
\newcommand{\ese}{\end{subequations}}
\newcommand{\beqa}{\begin{eqnarray}}
\newcommand{\eeqa}{\end{eqnarray}}
\newcommand{\beqar}{\begin{eqnarray*}}
\newcommand{\eeqar}{\end{eqnarray*}}
\newcommand{\bi}{\begin{itemize}}
\newcommand{\ei}{\end{itemize}}
\newcommand{\bn}{\begin{enumerate}}
\newcommand{\en}{\end{enumerate}}

\newcommand{\ba}{\begin{array}}
\newcommand{\ea}{\end{array}}
\newcommand{\bc}{\begin{center}}
\newcommand{\ec}{\end{center}}
\newcommand{\nnr}{\nonumber \\}

\definecolor{darkgreen}{rgb}{0,0.3,0}
\definecolor{mgreen}{rgb}{0,0.6,0}
\definecolor{darkblue}{rgb}{0,0,0.3}
\definecolor{darkred}{rgb}{0.7,0,0}

\begin{document}
\begin{titlepage}
\begin{flushright}\vspace{-3cm}
{\small
IPM/P-2015/nnn \\
\today }\end{flushright}
\vspace{0.5cm}
\begin{center}
\centerline{{\fontsize{18pt}{12pt}\selectfont{\bf{Black Rings in U(1)$^3$ Supergravity and their dual 2d CFT}}}} \vspace{10mm}
\centerline{{\Large{\bf{}}}} \vspace{8mm}
\centerline{\fontsize{15pt}{12pt}\selectfont{\bf{S. Sadeghian$^{a,b}$, H. Yavartanoo$^{c}$}}}
\vspace{5mm}
\normalsize
\bigskip\medskip
{$^a$ \it Department of Physics, Azzahra University,   Tehran 19938-91176, Iran}\\
\smallskip
  {$^b$ \it School of Physics, Institute for Research in Fundamental
 Sciences (IPM),\\ P.O.Box 19395-5531, Tehran, Iran}\\
\smallskip
 {$^c$ \it  State Key Laboratory of Theoretical Physics, Institute of Theoretical Physics,
 Chinese Academy of Sciences, Beijing 100190, China. }
\end{center}
\smallskip
\begin{abstract}
\noindent
\end{abstract}
We study the near horizon geometry of black ring solutions in five-dimensional U(1)$^3$
supergravity with three electric dipole charges and one angular momentum.
We consider the extremal vanishing horizon (EVH) limit of these solutions and show that the near horizon geometries develop
locally AdS$_3$ throats which at the near-EVH near horizon limit the  AdS$_3$ factor turns to a BTZ black hole.  By analysing the first law of thermodynamics for black rings we show that at EVH limit it reduces to the first law of thermodynamics for BTZ black holes. Using the AdS$_3$/CFT$_2$ duality, we propose a dual CFT 
to describe the near-horizon low energy dynamics of near-EVH black
rings.  We also discuss the connection between our CFT proposal
and the Kerr/CFT correspondence in the cases where these two overlap.

\end{titlepage}
\setcounter{footnote}{0}
\renewcommand{\baselinestretch}{1.05}  
\addtocontents{toc}{\protect\setcounter{tocdepth}{2}}
\tableofcontents

\section{Introduction}

One of the main challenges that a quantum theory of gravity should address is
determining  what degrees of freedom account for black hole's entropy.
There has been a lot of progress in understanding of black hole thermodynamics
in some special cases within string theory, involving extremal and near-extremal
black holes, following the eminent work by Strominger and Vafa \cite{Strominger:1996sh}. This
analysis has been generalised to many other extremal and near extremal black hole solutions in different theories of gravity in different dimensions.

If the near-horizon geometry of a black hole solution contains an AdS$_3$ factor, there exists a simpler microscopic model to describe black hole entropy. Since quantum gravity in asymptotically
AdS$_3$ space has a dual conformal field theory (2d-CFT), the entropy
of these extremal black holes can be counted, using the universal properties of the dual 2d-CFT \cite{Strominger:1997eq}.

More recently, there has been several attempts to find holographic description for more generic extremal black holes,  which have AdS$_2$ throats in their near horizons. These involved either the AdS$_2$/CFT$_1$ correspondence
\cite{Gupta:2008ki} or the Kerr/CFT correspondence \cite{Guica:2008mu}. For black holes with a compact horizon these proposals suggest the existence of a non-dynamical dual description. Thus finite energy excitations are not allowed and vacuum degeneracy of the dual CFT accounts for the macroscopic black hole entropy.

Here we continue our investigations of the Extremal Vanishing Horizon (EVH) black holes by studying five dimensional rotating black ring solutions to the U(1)$^3$ supergravity. These solutions have been found recently \cite{Lu:2014afa}.  They carry a single angular momentum, mass and three electric dipole charges. We show that the near horizon geometry of these solutions at the EVH limit contains an AdS$_3$ throat. The EVH limit can be achieved in two different ways, stationary and static limit. In the static limit, where the angular momentum vanishes, the AdS$_3$ part of the near horizon turns to a BTZ black hole at near-EVH limit. Using AdS/CFT correspondence, we propose a 2d CFT that governs near-EVH black ring dynamics and accounts the entropy. 

In the static EVH limit black rings where the angular momentum vanishes and we have static solutions, we find that near horizon geometry is well-defined in five dimensions. This is in contrast with the static EVH black  hole  solutions where the near horizon geometry is well-defined only after uplifting the solutions to higher dimensions \cite{Johnstone:2013eg}. However, we argue that the excitations do not lead to the BTZ in the near horizon geometry.

We also study the first law of thermodynamics for these black rings. We show that despite the existence of the conical singularity one can define thermodynamics quantities properly and they obey the fist law. This reduces to the first law of thermodynamics for the BTZ black hole when we take the near-EVH limit.

\section{Dipole charge black rings in five-dimensional U(1)$^3$ supergravity}
We consider a particular class of black ring solution to the  five-dimensional U(1)$^3$ supergravity.  We obtain the theory by dimensional reduction of eleven dimensional supergravity on a six dimensional space T$^2\times$T$^2\times$T$^2$, using following ansatz for metric and the form field
\bea
&& ds_{11}^2=ds_5^2+X_1^2(dz_1^2+dz_2^2)+X_2^2(dz_3^2+dz_4^2)+X_3^2(dz_5^2+dz_6^2), \\
&& 
\;\;\; \mathcal{A}=A_1 dz_1\wedge dz_2+A_2 dz_3\wedge dz_4+A_3 dz_5\wedge dz_6 ,
\eea
assuming that the scalar fields $X_i$ obey constraint $X_1X_2X_3=1$, which ensure that the internal 6-dimensional space has a fixed volume\footnote{By adding a warpped factor and allowing a non-compact internal space the same KK-reduction ansatz also works for the gauged supergravity\cite{Colgain:2014pha}.}. The bosonic  part of the reduced action is given by

\be
\label{sugra}
S=\frac{1}{16\pi G_5}\int dx^5 \left( R-\frac{1}{2} (\partial \vec{\Phi})^2-\frac{1}{4}\sum_{i=1}^3 X_i^{-2} F_i^2+\frac{1}{4}\epsilon_{\mu\nu\rho\sigma\lambda}F_1^{\mu\nu}F_2^{\rho\sigma}A_3^{\lambda}\right),
\ee
where we defined $F_i=dA_i$, $\vec{\Phi}=(\Phi_1,\Phi_2)$ and 
\be
X_1=e^{ -\frac{1}{\sqrt{2}}\Phi_1+\frac{1}{\sqrt{6} }\Phi_2},\quad X_2=e^{\frac{1}{\sqrt{2}}\Phi_1+\frac{1}{\sqrt{6} }\Phi_2},\quad  X_3=e^{-\frac{2}{\sqrt{6}}\Phi_2},
\ee
 
Alternatively, the five dimensional theory can be obtained from T$^4\times$S$^1$ compactification of IIB supergravity through the following ansatz
\bea
&& ds_{10}^2=X_3^{\frac{1}{2}} dx_5^2+X_3^{-\frac{3}{2}} (dz_5+A_3)^2+X_1X_3^{\frac{1}{2}}(dz_1^2+dz_2^2+dz_3^2+dz_4^2),\\
&& e^{2\phi}=\frac{X_1}{X_2},\quad\qquad  F_{(3)}=X_1^{-2}\star F_1 + F_2\wedge (dz_5+A_3),
\eea
where $\star F_1$ is the dual form with respect to the five-dimensional metric, $F_{(3)}$ is the RR 3-form field strength and $\phi$ is the dilaton. 
\subsection{Neutral single-spin black ring solution}
Let us start with the simplest black ring solution in five dimensions, i.e neutral single-spin black ring solution.  The solution has been constructed in \cite{Emparan:2001wn}
. We adapt the notation \cite{Emparan:2006mm} which is more suitable for our study\footnote{To compare this solution to the general one discussed in \cite{Chen:2011jb} we take $\sigma\to 0, \mu\to\nu, \xi\to\lambda, k\to\sqrt{\frac{1-\lambda}{2(1-\nu^2) }} R$.}
\bea
\label{BRmetric}
&&\hspace{-11mm} ds^2=-\frac{F(y)}{F(x)}\left(dt-\frac{RC(1+y)}{\sqrt{K}F(y)}d\psi\right)^2\nnr
&&+\; \frac{R^2 F(x)}{(x-y)^2}\left[-\frac{G(y)}{K F(y)} d\psi^2-\frac{dy^2}{G(y)}+\frac{dx^2}{G(x)}+\frac{G(x)}{ K F(x)}d\phi^2 \right], \eea
where 
\be
F(\xi)=1+\lambda \xi,\quad G(\xi)=(1-\xi^2)(1+\nu\xi),\quad   C=\sqrt{\lambda(\lambda-\nu)\frac{1+\lambda}{1-\lambda}},\quad K=\frac{(1-\nu)^2}{1-\lambda}\; .   
\ee
The dimensionless parameters $\nu$ and $\lambda$ lie in the range
\be
0<\nu\leq \lambda<1\;,
\ee
 and $ x\in [-1,1], \;\; y\in (-\infty,-1]$ and $\phi, \psi \in [0,2\pi] $.  Asymptotic infinity, which is a flat five dimensional Minkowski space-time is located at $x= y=-1$ and black ring horizon at $y=-\nu^{-1}$. The solution has a conical singularity at $x=1$. The deficit/excess is given by
 \be
 \Delta=2\pi \left(1-\frac{(1+\nu)\sqrt{1-\lambda}}{(1-\nu)\sqrt{1+\lambda}}\right)\;.
 \ee
 To avoid the conical singularity ($\Delta=0$) we must take
 \be
 \label{NCS0}
 \lambda=\lambda_c=\frac{2\nu}{1+\nu^2}\;.
 \ee

In our analysis we relax this condition.  The ADM mass, angular momentum, angular velocity at the horizon, entropy and Hawking temperature  for these solutions are given by
\bea
\label{Thermo1}
&& M_0=\frac{3\pi R^2}{4G_5}\frac{\lambda}{1-\nu},\quad J_{\psi}=\frac{\pi R^3}{2G_5}\frac{\sqrt{\lambda(\lambda -\nu)(1+\lambda)}}{(1-\nu)^2}\quad \Omega_{\psi}=\frac{1}{R}\sqrt{\frac{\lambda-\nu}{\lambda(1+\lambda)}},\nnr
&& S=\frac{2\pi^2 R^3}{G_5}\frac{\nu^{3/2}\sqrt{\lambda (1-\lambda^2)}}{(1-\nu)^2(1+\nu)},\quad T_H=\frac{1+\nu}{4\pi R}\sqrt{\frac{1-\lambda}{{\lambda\nu(1+\lambda)}}}\;.
\eea
It is straightforward to check that for general values of the parameters we have the Smarr relation
\be
M_0=\frac{3}{2}\left(TS+\Omega_{\psi} J_{\psi}\right)\;.
\ee

When the conical singularity is absent $\lambda=\lambda_c$, the first law of thermodynamics 
\be
dM_0=TdS+\Omega_{\psi}dJ_{\psi},
\ee
is obeyed. This is no longer valid for unbalanced black ring with $\lambda \neq \lambda_c$. This is due to the energy corresponding to the conical singularity in space-time, which needs to be considered in the first law properly.

\subsubsection{First law of thermodynamics in the presence of conical singularity}

Although it seems that the existence of  conical singularity is a pathology for defining thermodynamics of the black ring,  the solutions have well defined Euclidean action, and therefore we expect they have well defined thermodynamic properties. In \cite{Herdeiro:2009vd}
it is argued that for asymptotic flat solutions with conical singularity if we work with the appropriate
set of thermodynamic variables, the area relation of Bekenstein-Hawking entropy still holds, but the mass which enters in the first law of thermodynamics is different from the ADM mass and the difference is the energy associated with the
conical singularity, as seen by an asymptotic, static observer. The first law of thermodynamics should be modified accordingly,
\be
\label{firstlaw2a}
dM=T_HdS+\Omega_I dJ_I + \mathcal{P} d\mathcal{A}\; .
\ee
The first two terms on the right hand side are the standard ones, $J_I$ indicates charge/angular momentom of the black hole and $\Omega_I$ is its momentum conjugate. The last term accounting for the effect of the conical singularity, which is exerting a pressure $\mathcal{P}$ with world-volume spanning a space-time area $\beta \mathcal{A}$, computed in the
Euclidean section where $\beta$ is the periodicity of the Euclidean time. 

The energy associated with the conical singularity which contributes to the mass defined by (\ref{firstlaw2a}) can be evaluated and is given by $E_{int}\equiv M_0-M=-\mathcal{P} \mathcal{A}$. 

For solution (\ref{BRmetric}) we obtain
\be
{\mathcal A}=\frac{\pi R^2 \sqrt{1-\lambda^2}}{4G_5(1+\nu)},\qquad {\mathcal P}=\frac{(1+\nu)\sqrt{1-\lambda}}{(1-\nu)\sqrt{1+\lambda}}-1\;.
\ee
Modified Smarr formula is given by
\be
M=\frac{3}{2}\left(TS+\Omega_{\psi} J_{\psi}\right) +{\mathcal{P A}}\;.
\ee

\subsubsection{Myers-Perry black hole limit}

As a particular limit of metric (\ref{BRmetric}) we can obtain the Myers-Perry black hole with rotation in a single plane if $R\to 0$ and $\lambda, \nu \to 1$, while maintaining
fixed the parameters $a, m$. This is achieved by taking the limit
\be
\lambda=1-\epsilon^2 , \quad \nu =1-\frac{m\epsilon^2}{m-a^2} ,\quad R=\frac{m\epsilon}{\sqrt{2(m-a^2)}} ,\qquad \epsilon\to 0\;.
 \ee
 Note that in this limit, condition (\ref{NCS0}) is satisfied and the solution does not have conical singularity while ADM mass, Hawking temperature, entropy and angular momentum remain finite,
 \be
 M_0=\frac{3}{8}\pi m,\quad T=\frac{\sqrt{m-a^2}}{2\pi m},\quad J=\frac{1}{4}\pi a m,\quad S=\frac{1}{2}\pi^2 m\sqrt{m-a^2}\;.
 \ee

It is more convenient to use  following coordinate transformation
 \be
 \label{CT1}
 x=-1+\frac{(m-a^2)  \cos^2\theta}{r^2-(m-a^2)\cos^2\theta}\;\epsilon^2,\quad y=-1-\frac{(m-a^2)  \sin^2\theta}{r^2-(m-a^2)\cos^2\theta}\;\epsilon^2\;,
 \ee
 leading to
 \be
 ds^2=-dt^2+\frac{m}{\Sigma}(dt-a^2 \sin^2\psi)^2+\frac{\Sigma}{\Gamma}dr^2+\Sigma d\theta^2+(a^2+r^2)\sin^2\theta d\psi^2 +r^2\cos^2\theta d\phi^2\;,
 \ee
 where
 \be
 \Sigma= r^2+a^2\cos^2\theta,\quad \Gamma=r^2-(m-a^2)\;.
 \ee
The coordinate transformation (\ref{CT1}) looks like a small expansion near the asymptotic infinity which is located at $x=y=-1$. In this limit, the horizon which is located at $y=-\nu^{-1}  \simeq -1-\epsilon^2$, may also seem close to the asymptotic infinity. But we must note that in this limit $xx$ and $yy$ components of the metric are also order of $\epsilon^2$ and one can show that the horizon is infinitely far from the asymptotic infinity. Moreover for a generic value of $r$, points in (\ref{CT1}) are at finite distance from the horizon and arbitrarly far from the asymptotic infinity. It is also useful to compare  (\ref{CT1}) with taking the asymptotic limit of black ring solutions which is obtained by
\be
x=-1+\epsilon^n \tilde{x}, \quad y=-1-\epsilon^n \tilde{y},\quad n>2,
\ee
and yields 
\be
\label{FSM}
ds^2=-dt^2+\frac{m\; \epsilon^{2-n}}{(\tilde{x} +\tilde{y})^2}\left[ \frac{d\tilde{x}^2}{\tilde{x}}+ \frac{d\tilde{y}^2}{\tilde{y}} +\tilde{x}d\phi^2 +\tilde{y}d\psi^2 \right]\;.
\ee
This is a five dimensional flat space-time metric, written in the ring coordinate system.  
\subsubsection{ The EVH limit}

For the black rings (\ref{BRmetric}) with finite mass, the extremal limit where Hawking temperature vanishes corresponds to $\lambda\to 1$. However from (\ref{Thermo1}), in this limit entropy vanishes while the ADM mass and angular momentum remain finite with $M^3=\frac{27}{32}\pi J^2$. 
   
The extremal vanishing horizon (EVH) limit for many different of black hole/ring solutions in different dimensions is studied in detail (see e.g \cite{Fareghbal:2008ar}-\cite{Sadeghian:2014tsa}). In \cite{Sadeghian:2014tsa}, we have shown that under very general assumptions the near horizon geometry of an EVH black hole develops an AdS$_3$ factor. This AdS$_3$ is replaced by a BTZ black hole when we take the near horizon limit of a near-EVH black hole.

Taking $\lambda=1-\epsilon^2$ and expanding thermodynamic quantities for small $\epsilon$ we find that $T\sim S\sim \epsilon$. Then, we are interested in studying the near horizon geometry of the black ring solution in this limit. By taking the near horizon limit we want to focus on the region of space-time close to the horizon.  Since we introduce a small parameter $\epsilon$, while we take near horizon geometry one can imagine three possibilities: where distance to the horizon is small but much larger than $\epsilon$, is of the same order, or much smaller than $\epsilon$. As we have shown in \cite{Sadeghian:2014tsa}, the first two possibilities correspond to the near-horizon EVH (NHEVH) geometry and near-EVH near-horizon limit, respectively. The last case would correspond to far from EVH case and we will exclude it.

Finding the near horizon as defined above is subtle in the original ring coordinate system. It is then useful to consider the following coordinate transformation
\be\label{CTn}
x=-1+\frac{(1-\nu)(1-\lambda)\cos^2\theta}{r^2+ \nu(1-\lambda)\cos^2\theta},\quad y=-1-\frac{(1-\nu)(1-\lambda)\sin^2\theta}{r^2+ \nu(1-\lambda)\cos^2\theta},
\ee   
where $\theta\in [0,\pi/2]$. In this coordinate system the horizon is located at $r=0$ and asymptotic infinity at $r=\infty$. Note that this coordinate system covers region only outside of the horizon. Now to take near horizon geometry we scale $r\to\delta r$ with $\epsilon\ll\delta\ll1$ and expanding the metric for small parameters $\delta$ and $\epsilon$. This ensures we are zooming into the near horizon region of distance about $\delta$. By expanding the metric  we get
\bea
\label{NHG2}
&&ds^2= - 4{r^2\cos^2\theta}\delta^2 dt^2 +\frac{2R^2\cos^2\theta}{1-\nu} dr^2 +\frac{2R^2\cos^2\theta}{1-\nu}\delta^2 d\phi^2 \nnr
&&\hspace{11mm} +\frac{2R^2\cos^2\theta }{1-\nu} d\theta^2 +\frac{R^2\sin^2\theta}{2(1-\nu)\cos^2\theta}\left(d\psi-\frac{\sqrt{2(1-\nu)}}{R}dt\right)^2\;.
\eea
The first line in above metric has an AdS$_3$ factor, anticipated by theorems 1 and 2 in \cite{Sadeghian:2014tsa}. It is more convenient to use the  following coordinate transformation
\bea
d\phi=\frac{d\tilde{\phi}}{{\delta} },\quad  d\psi=2d\tilde{\psi}+\frac{\sqrt{2(1-\nu)}}{2R} dt   ,\quad dt=\frac{R }{\sqrt{2-2\nu}} \frac{ d\tau}{\delta}\;.
\eea
Metric (\ref{NHG2}) then turns into
\be
ds^2=\frac{2R^2}{1-\nu}\cos^2\theta\left[-r^2d\tau^2+\frac{dr^2}{r^2}+r^2d\tilde{\phi}^2\right]+ \frac{2R^2}{1-\nu}\left(\cos^2\theta d\theta^2+\frac{\sin^2\theta}{\cos^2\theta}d\tilde{\psi}^2\right)\;.
\ee 
The AdS$_3$ angular direction $\tilde{\phi}$ is pinching with range $\tilde{\phi}\in[0,{2\pi}{\delta}]$. It is also clear that the AdS$_3$ in the near horizon does not turn to a BTZ at near EVH limit. This can be understood by looking at the more general black ring solution with two angular momenta and its EVH limit \cite{Ghodsi:2013soa}. The only possible excitations are in the second angular momentum direction. 

\subsection{Adding dipole charges}
The black ring solutions which we are studying here have mass, three dipole charges and one angular momentum. These solutions were constructed in \cite{Lu:2014afa}.  A more general class of black ring solutions with two independent angular momenta and electric dipole charges was constructed in \cite{Feldman:2014wxa}
\bea
\label{solution}
&&ds_5^2=- \frac{y}{xU^{{1\over 3}}}\left[dt+({y^{-1}}-{\eta_3^{-2}}){aQ_4}\;d\phi_1  \right]^2 +  \frac{x}{y (x-y)^2 U^{{1\over 3}}}\left(-a^2G(y)d\phi_1^2-\frac{\mathcal{H}(y)}{G(y)}dy^2 \right) \cr\cr
&&\hspace{12mm} +\frac{U^{2\over 3}}{(x-y)^2}\left(\frac{\mathcal{H}(x)}{G(x)}dx^2+a^2G(x)d\phi_2^2 \right),
\eea

where $x\in [\eta_3^2, \eta_4^2],  y\in [-\eta_1^2, \eta_3^2]$ and we defined 
\bea
\label{functions}
&&  {\mathcal H}(\xi)=\xi h_1(\xi)  h_2(\xi)  h_3(\xi),\quad  G(\xi)= -\mu^2(\xi+\eta_1^2)(\xi+\eta_2^2)(\xi-\eta_3^2)(\xi-\eta_4^2) ,      
   \nnr
&&a=\frac{ 2\eta_3 \sqrt{(\eta_3^2+q_1)(\eta_3^2+q_2)(\eta_3^2+q_3)}}{\mu^2 (\eta_1^2+\eta_3^2) (\eta_2^2+\eta_3^2) (\eta_4^2-\eta_3^2)},   \;\; Q_i^2=-G(-q_i),  \quad Q_4^2=\mu^2 \eta_1^2 \eta_2^2 \eta_3^2 \eta_4^2,
\nnr
&&      h_i(\xi)\equiv \xi+q_i ,\qquad U_i\equiv \frac{h_i(y)}{h_i(x)},\qquad U\equiv U_1 U_2 U_3,\quad i=1,2,3, \nnr
&& \eta_1>\eta_2>0, \qquad \eta_4>\eta_3>0\;.
\eea


Scalar and gauge fields are
\bea
X_i=\frac{U_i}{U^{1\over 3}},\qquad A_i=\frac{aQ_i}{h_i(x)}d\phi_2.
\eea 
It is straightforward to check that the black ring solution (\ref{solution}) for a generic value of parameters has a conical singularity.  However, as it has been shown in \cite{Lu:2014afa} this can be avoided by imposing the following constraint on the black ring parameter space

\be
\label{no-singularity-condition}
\frac{(\eta_1^2+\eta_3^2)(\eta_2^2+\eta_3^2)\sqrt{ {\mathcal H}(\eta_4^2)}}{(\eta_1^2+\eta_4^2)(\eta_2^2+\eta_4^2)\sqrt{ {\mathcal H}(\eta_3^2)}}=1\;.
\ee
The neutral black ring solution is obtained in the $q_i\to\eta_1^2$ limit. To get metric (\ref{BRmetric}) one should perform following coordinate transformations 
\be
x\to \frac{c_1+\eta_1^2 x}{c_2-x},\quad y\to \frac{c_1+\eta_1^2 y}{c_2-y},\quad  c_1=\frac{\eta_1^2\eta_3^2+\eta_1^2\eta_4^2+2\eta_3^2\eta_4^2}{\eta_4^2-\eta_3^2},\quad c_2=\frac{2\eta_1^2+\eta_3^2+\eta_4^2}{\eta_4^2-\eta_3^2},
\ee
and the neural black ring parameters are given by
\bea\label{3parameters}
&& \lambda=\frac{\eta_1^2(\eta_4^2-\eta_3^2)}{\eta_1^2\eta_3^2+\eta_1^2\eta_4^2+2\eta_3^2\eta_4^2},\; \quad\nu=\frac{(\eta_1^2-\eta_2^2)(\eta_4^2-\eta_3^2)}{\eta_1^2\eta_2^2+\eta_3^2\eta_4^2+\sum_{j>i}\eta_i^2\eta_j^2},\nnr &&R^2=\frac{4(\eta_1^2+\eta_3^2)(\eta_1^2+\eta_4^2)}{\mu^2(\eta_4^2-\eta_3^2)(\eta_1^2\eta_2^2+\eta_3^2\eta_4^2+\sum_{j>i}\eta_i^2\eta_j^2)}\;.
\eea

 The solution (\ref{solution}) is over parametrised. This can be also seen in (\ref{3parameters}).  Among five parameters $\eta_i$ and $\mu$ only three combinations in (\ref{3parameters}) are independent which along with three parameters $q_i$ give the six independent parameters of the original dipole black ring. The absence of conical singularity ( \ref{no-singularity-condition}) eliminates one extra parameter and leaves five real parameters corresponding to three dipole charges, mass and angular momentum.

\subsubsection{Charges and thermodynamics}
The angular momentum and three dipole electric charges of the family of black ring solutions can be
evaluated using Komar and Gaussian integrals respectively,
\be\label{J}
J\equiv J_{\phi_1} =\frac{ \pi aQ_4}{4 G_5b^2\eta_3^4},\quad J_{\phi_2}=0\qquad D_i=\frac{\pi a Q_i}{4 G_5}\left(\frac{1}{q_i+\eta_3^2}-\frac{1}{q_i+\eta_4^2} \right)\;,
\ee
where
\be
b^2=\frac{\mu^2(\eta_1^2+\eta_3^2)(\eta_2^2+\eta_3^2)(\eta_4^2-\eta_3^2)}{4\eta_3^2(\eta_3^2+q_1)(\eta_3^2+q_2)(\eta_3^2+q_3)}\; .
\ee
The horizon is located at  $y=-\eta_2^2$ and its topology is S$^1\times$S$^2$ where S$^1$ is along the $\phi_1$ direction and S$^2$  corresponds to $(x,\phi_2)$ directions. The horizon structure determines the
thermodynamic properties of the black ring. The Hawking temperature can be computed through the
 surface gravity at the horizon, leading to
\be\label{Temp}
T_H =\frac{\mu \eta_3{(\eta_1^2-\eta_2^2)}(\eta_2^2+\eta_4^2)}{4\pi \eta_1 \eta_4 \sqrt{(q_1-\eta_2^2)(q_2-\eta_2^2)(q_3-\eta_2^2)}}\;,
\ee
and the Bekenstein-Hawking entropy, which is proportional to the area of the black ring horizon, is given by
\be 
\label{Entropy}
S=\frac{\pi^2 a^2 Q_4 (\eta_4^2-\eta_3^2)}{G_5\; \eta_2\eta_3^2 (\eta_2^2+\eta_4^2)}  \sqrt{(q_1-\eta_2^2) (q_2-\eta_2^2)(q_3-\eta_2^2)}\;.
\ee
The Killing horizon is generated by the Killing vector field $\ell =\partial_t+\Omega \partial_{\phi_1}$
 where $\Omega$ stands for the angular velocity on the horizon,
 \be\label{om}
 \Omega \equiv \Omega_{\phi_1}=\frac{\eta_2^2\eta_3^2}{a Q_4 (\eta_2^2+\eta_3^2) }\;.
 \ee

 The asymptotic region is located at $x=y=\eta_3^2$, where to get the Minkowski metric we make the coordinate transformation
 \be
r\sin\theta= \frac{\sqrt{\eta_3^2-y}}{b(x-y)},\quad r\cos\theta= \frac{\sqrt{x-\eta_3^2}}{b(x-y)}\;,
 \ee
 then taking limit $r\rightarrow \infty$ we get
 \be
 ds^2=-dt^2+dr^2+r^2(d\theta^2+\sin^2\theta d\phi_1^2 +\cos^2\theta d\phi_2^2)\;.
 \ee
 The ADM mass can be read off from the asymptotic falloffs of the metric
  
\be 
\label{M0}
M_0=\frac{\pi}{8 b^2 G_5}\left(\frac{3}{\eta_3^2}-\frac{1}{\eta_3^2+q_1}-\frac{1}{\eta_3^2+q_2}-\frac{1}{\eta_3^2+q_3} \right)\;.
\ee
The electrostatic potentials $\Phi_i$ associated with the electric dipole charges are given by
\be
\Phi_{D_i}=aQ_i\left(\frac{1}{q_i-\eta_2^2}-\frac{1}{q_i+\eta_3^2}\right)\;.
\ee

One can check that the ADM mass defined by (\ref{M0}) obeys the Smarr formula,
\be
\label{smarr}
M_0=\frac{3}{2}TS +\frac{3}{2}\Omega_1 J_{\phi_1} +\frac{1}{2}\sum_{i=1}^3\Phi_{D_i} D_i \;.
\ee

It is noted in \cite{Lu:2014afa} that when the condition (\ref{no-singularity-condition}) is satisfied and therefore solutions do not have the conical singularity, thermodynamics quantities defined above satisfy the first law of thermodynamics,
\be
\label{firstlaw1}
d M_0=T_HdS+\Omega dJ +\sum_{i=1}^3\Phi_{D_i}dD_i\;.
\ee 
However, this is not valid for generic value of charges where the solutions (\ref{solution}) have conical singularity.

For the black ring solution (\ref{solution}) the first law of thermodynamics reads
\be
 \label{firstlaw2}
d {M}=T_HdS+\Omega dJ +\sum_{i=1}^3\Phi_{D_i}dD_i + \mathcal{P} d\mathcal{A},
\ee
where ${\mathcal P}$ and  ${\mathcal A}$ are given by
\be
{\mathcal P}=-\frac{\Delta}{2\pi},\quad {\mathcal A}= \frac{{\mathrm{Area}}}{4G_5} \times {T_H}\;.
\ee
For solution (\ref{solution}), $\mathcal{P}$  is given by
\be
\mathcal{P}= \frac{ (\eta_1^2+\eta_4^2)(\eta_2^2+\eta_4^2)\sqrt{{\mathcal H}(\eta_3^2)} }{(\eta_1^2+\eta_3^2)(\eta_2^2+\eta_3^2)\sqrt{{\mathcal H}(\eta_4^2)}}-1\;.
\ee
The world volume of the conical defect is a 3d surface and its area is given by
\be
 {\mathrm{Area}} =\int \sqrt{{\mathrm{det}} h} \; d\tau d\phi_1 dy ,
\ee
where $h$ is induced metric on the 3d surface spanned by the conical singularity for the solution (\ref{solution}). It reads
\be
 {\mathcal A}=\frac{\pi \eta_3\eta_4\sqrt{{\mathcal H}(\eta_3^2) {\mathcal H}(\eta_4^2)}}{G_5\mu^2 (\eta_1^2+\eta_3^2)(\eta_2^2+\eta_4^2)(\eta_4^2-\eta_3^2)^2}\;.
\ee
From (\ref{firstlaw2}) we find that
\be
\label{mass-correction}
E_{int}=\frac{\pi \sqrt{{\mathcal H}(\eta_3^2)}\left[    (\eta_1^2+\eta_3^2)(\eta_2^2+\eta_3^2) \sqrt{{\mathcal H}(\eta_4^2)}      - (\eta_1^2+\eta_4^2)(\eta_2^2+\eta_4^2) \sqrt{{\mathcal H}(\eta_3^2)}\right] }{G_5 \mu^2  (\eta_1^2+\eta_3^2)^2(\eta_2^2+\eta_3^2)(\eta_2^2+\eta_4^2)(\eta_4^2-\eta_3^2)^2 }\;.
\ee
From (\ref{no-singularity-condition}) it is clear that in the absence of conical singularity the last term in the first law of thermodynamics (\ref{firstlaw2}) and the correction to black ring mass (\ref{mass-correction}) vanish and we get back to the familiar form of the first law\footnote{In \cite{Lu:2014afa} authors noticed that the black ring solutions obey the first law (\ref{firstlaw1}) only in the absence of conical singularity. }.  One can check these satisfy modified Smarr mass formula
\be
{M}=\frac{3}{2}TS +\frac{3}{2}\Omega_1 J_{\phi_1} +\frac{1}{2}\sum_{i=1}^3\Phi_{D_i} D_i +\mathcal{P} \mathcal{A}\;.
\ee
Using the fact that $E_{int}=-\mathcal{PA}$, this reduces to (\ref{smarr}). 

\subsubsection{Near horizon geometry and Kerr/CFT correspondence}
The extremal black ring solutions correspond to  $\eta_1=\eta_2$ limit where Hawking temperature vanishes. To take the near horizon geometry, we consider the following rescalings 
\bea
&&\hspace{-10mm} y=-\eta_2^2+\epsilon Y,\quad\qquad t=k\frac{\tau}{\epsilon},\quad\qquad \phi_1=\tilde{\phi}_1-\omega  t, \nnr
&& \hspace{-10mm}k=\frac{\eta_2\eta_4 \sqrt{(q_1-\eta_2^2)(q_2-\eta_2^2)(q_3-\eta_2^2)}}{\mu\eta_3(\eta_2^2+\eta_4^2)},\quad \omega=\frac{\mu(\eta_2^2\eta_3^2+\eta_3^4-\eta_3^2\eta_4^2-\eta_2^2\eta_4^2)}{2\eta_4\sqrt{(q_1+\eta_3^2)(q_2+\eta_3^2)(q_3+\eta_3^2)}},
\eea
and take limit $\epsilon\rightarrow 0$. We get
\be
ds^2=A(x)\left(-Y^2d\tau^2+\frac{dY^2}{Y^2}+B(x)\left(d\tilde{\phi}_1+k_{\tilde{\phi}_1} Y d\tau\right)^2\right) + C(x) \left(dx^2 + E(x) d\phi_2^2\right)\,.
\ee
The three dimensional part of the near horizon metric is a U(1) bundle over AdS$_2$. Functions $A$ and $B$ and constant $k_{\phi_1}$ are given by
\bea
&&k_{\tilde{\phi}_1}=\frac{\eta_3^2 (\eta_4^2-\eta_3^2)\sqrt{(q_1-\eta_2^2)(q_2-\eta_2^2)(q_3-\eta_2^2)}}{2\eta_2 (\eta_2^2+\eta_4^2) \sqrt{\mathcal{H}(\eta_3^2)}},
\nnr && A(x)=\frac{((q_1-\eta_2^2)(q_2-\eta_2^2)(q_3-\eta_2^2))^{2\over3}x^{2\over3}\mathcal{H}^{1\over3}(x)}{\mu^2(\eta_2^2+\eta_3^2)(\eta_2^2+\eta_4^2)(x+\eta_2^2)^2},\nnr
&&B(x)=\frac{4\eta_2 \eta_4^2 (\eta_2^2+\eta_4^2)^2\mathcal{H}(\eta_3^2) (x+\eta_2^2)^2}{\eta_3^2 (q_1-\eta_2^2)(q_2-\eta_2^2)(q_3-\eta_2^2)(\eta_2^2+\eta_3^2)(\eta_2^2+\eta_4^2)(\eta_4^2-\eta_3^2)^2 x^2}\; ,
\nnr
&& C(x)=\frac{\left((q_1-\eta_2^2)(q_2-\eta_2^2)(q_3-\eta_2^2)\right)^{2\over3}x^{2\over3}\mathcal{H}^{1\over3}(x)}{\mu^2 (\eta_4^2-x)(x-\eta_3^2)(x+\eta_2^2)^4},
\nnr
&&E(x)=\frac{4(\eta_4^2-x)^2(x-\eta_3^2)^2(x+\eta_2^2)^4 \mathcal{H}(\eta_3^2)}{(\eta_2^2+\eta_3^2)^4(\eta_4^2-\eta_3^2)^2 \mathcal{H}(x)}\;.
\eea

According to the Kerr/CFT dictionary, this fixes the dual CFT Frolov-Thorne temperature
\be
T_{\tilde{\phi}_1}=\frac{1}{2\pi k_{\tilde{\phi}_1}},\;
\ee
The central charge of corresponding chiral Virasoro algebra is obtained from the asymptotic
symmetry group analysis 
\be\label{KC}
c_{\tilde{\phi}_1}=\frac{12\pi \eta_3\eta_4 (q_1-\eta_2^2)(q_2-\eta_2^2)(q_3-\eta_2^2)\sqrt{\mathcal{H}(\eta_3^2)}}{G_5 \mu^3 (\eta_2^2+\eta_3^2)^4(\eta_2^2+\eta_4^2)^2},
\ee
and entropy of the extremal black ring can be reproduced upon using Cardy's formula
\be
S=\frac{\pi^2}{3}c_{\tilde{\phi}_1}T_{\tilde{\phi}_1}=\frac{4\pi^2\eta_2\eta_4\mathcal{H}(\eta_3^2)\sqrt{(q_1-\eta_2^2)(q_2-\eta_2^2)(q_3-\eta_2^2)}}{G_5 \mu^3\eta_3(\eta_4^2-\eta_3^2)(\eta_2^2+\eta_4^2)(\eta_2^2+\eta_3^2)^4},
\ee
which is equal to the Bekenstein-Hawking entropy (\ref{Entropy}) evaluated at the extremal point.

\section{Extremal Vanishing Horizon (EVH) limits of dipole black ring}

The extremal limit where Hawking temperature (\ref{Temp}) vanishes corresponds to $ \eta_1=\eta_2$ or $\eta_3=0$. In the space of extremal black ring solutions we are interested in studying those with vanishing horizon area. We are physically defining the subset of EVH configurations as a limit of near-extremal black rings in which area of horizon $A_h \sim T_H \sim \epsilon \rightarrow 0$, keeping the ration $A_h/T_H$ finite. Inspection of (\ref{Temp}) and (\ref{Entropy}) reveals $\eta_1 \sim \eta_2 \sim \epsilon$ or $\eta_3\sim \epsilon$ as the EVH regions. 
\subsection{Stationary Case : $\eta_3\sim 0$}
Taking vanishing limit of $\eta_3$ while keeping other black ring parameters finite, we get the entropy and Hawking temperature vanishing at the same rate.  Dipole charges are all vanishing in this limit while angular momentum remains finite. This branch is rather similar to the EVH limit of neutral single spin black ring studied in section (2.1).  The shrinking cycle of horizon is $\phi_2$ which is inside S$^2$ part of horizon.  To take near horizon limit we consider coordinate transformation (\ref{CTn}) where $\lambda, \nu$ and $R$ are defined by (\ref{3parameters}) and then scale $r\to \delta r$ with $\epsilon \ll \delta \ll 1$. Expanding the metric for small parameters gives
\be\label{ads3eta3}
ds^2=\frac{4q_1q_2q_3}{\mu^2\eta_1^2\eta_2^2\eta_4^2}\cos^2\theta\bigg[-\rho^2d\tau^2 +\frac{d\rho^2}{\rho^2}+\rho^2d\tilde{\phi}_2^2\bigg]+ \frac{4q_1q_2q_3}{\mu^2\eta_1^2\eta_2^2\eta_4^2}\left(\cos^2\theta d\theta^2 +\frac{\sin^2\theta}{\cos^2\theta} d\tilde{\phi}_1^2\right),
\ee
where 
\be
\rho=\sqrt{\frac{2\eta_2^2(\eta_1^2+\eta_4^2)}{2\eta_1^2\eta_2^2+\eta_1^2\eta_4^2+\eta_2^2\eta_4^2} } \;\; r, \quad \tilde{\phi}_2=\delta\phi_2,\quad \tau=\frac{\mu\eta_1\eta_2\eta_4}{\sqrt{4q_1q_2q_3}}\delta t,\quad \tilde{\phi}_1=\phi_1-\frac{\mu\eta_1\eta_2\eta_4}{\sqrt{4q_1q_2q_3}} t\;.
\ee
The first three terms in metric (\ref{ads3eta3}) forms a pinching AdS$_3$ space. Similarly to what we discussed in section (2.1), this cannot be excited to BTZ. 

\subsection{Static Case: $\eta_1\sim\eta_2\sim 0$}
Assuming $\eta_{1,2}$ are small and same order of $\epsilon$ and keeping the other black ring parameters finite. We obtain 
\be \label{temp-ent}
T_H=\frac{\mu \eta_3\eta_4 (\alpha_1^2-\alpha_2^2)}{\pi \alpha_1 \sqrt{q_1q_2q_3}}\;\epsilon +{\mathcal O}(\epsilon^3),\qquad S=\frac{4\pi^2 {\mathcal{H}}(\eta_3^2)\sqrt{q_1q_2q_3}\;\alpha_1 }{\mu^3\eta_3^9\eta_4(\eta_4^2-\eta_3^2)G_5}\;\epsilon +{\mathcal O}(\epsilon^3)\;.
\ee
The mass and electric dipole charges of black rings remain finite while the angular momentum vanishes at this limit $J\sim \epsilon^2$,  in contrast to the stationary case. This is static limit of solution (\ref{solution}). It also worths to note that, the cycle which its shrinking causes vanishing of horizon area is the S$^1$ part of the horizon while the size of S$^2$ part  remains finite.

\subsubsection{Near Horizon Geometry Analysis}
We study the near horizon geometries corresponding to the 
EVH black rings together with their near-extremal versions. Assuming parameters $\eta_{1}\sim\eta_2\sim \epsilon$ are small and studying deep interior geometry of the EVH black rings by
expanding in small $\delta$ for $y = \delta^2 Y^2$ where $\epsilon \ll \delta$, leading terms in the metric expansion are
\bea
\label{Shrink}
&& ds^2=\frac{4(q_1q_2q_3 )^{2\over 3} {\mathcal H}^{1\over 3}(x)}{ \mu^2\eta_3^2\eta_4^2 x^{4\over 3}}\left[-\frac{\epsilon Y^2 dt^2}{C^2K^2}+\frac{dY^2}{Y^2}+\frac{\epsilon Y^2 {d{\phi}}_1^2}{ C^{2}} \right] +
\frac{ (q_1q_2q_3 )^{2\over 3}{\mathcal H}(x)^{1\over 3}}{ \mu^2 x^{10\over 3} (x-\eta_3^2)(\eta_4^2-x)} \bigg(dx^2 
\nonumber \\
&& \hspace{10mm}
+{\frac{4  {\mathcal H}(\eta_3^2) \; x^4(x-\eta_3^2)^2(\eta_4^2-x)^2}{\eta_3^8(\eta_4^2-\eta_3^2)^2{\mathcal H}(x)}}d\phi_2^2\bigg),
\eea
where 
\be
 C^2=\frac{\eta_3^2 q_1q_2q_3 (\eta_4^2-\eta_3^2)^2}{\eta_4^2{\mathcal H}(\eta_3^2)},\quad K^2=\frac{4\eta_4^2{\mathcal H}(\eta_3^2)}{\mu^2\eta_3^6(\eta_4^2-\eta_3^2)^2}\;.
\ee
Extremality condition determines the scaling $\epsilon Y^2 dt^2$ together with $\frac{dY^2}{Y^2}$ giving rise to an AdS$_2$
throat responsible for the SO(2, 1) isometry enhancement of the near horizon geometry of black rings. However the new feature here is the vanishing size of the one-cycle along $\phi_1$ direction as $\epsilon Y^2$. This is responsible
for the vanishing of the entropy and transforms the standard AdS$_2$ throat into a local $pinching$ AdS$_3$
throat.  Thus the near horizon geometry  is obtained by considering the limit
\be
\eta_{1,2}=\alpha_{1,2} \epsilon,\quad Y=C \rho ,\quad     dt=\frac{{K}d\tau}{ \sqrt{\epsilon}}, \quad \phi_1=\frac{\psi}{\sqrt{\epsilon}} \;, 
\ee
The resulting metric is

\bea\label{AdS3}
&& ds^2=\frac{4(q_1q_2q_3 )^{2\over 3} {\mathcal H}^{1\over 3}(x)}{ \mu^2\eta_3^2\eta_4^2 x^{4\over 3}}\left[-{\rho^2 d\tau^2}+\frac{d\rho^2}{\rho^2}+{\rho^2 {d\psi}^2} \right] +
\frac{ (q_1q_2q_3 )^{2\over 3}{\mathcal H}(x)^{1\over 3}}{ \mu^2 x^{10\over 3} (x-\eta_3^2)(\eta_4^2-x)} \bigg(dx^2 
\nonumber \\
&& \hspace{10mm}
+{\frac{4  {\mathcal H}(\eta_3^2) \; x^4(x-\eta_3^2)^2(\eta_4^2-x)^2}{\eta_3^8(\eta_4^2-\eta_3^2)^2{\mathcal H}(x)}}d\phi_2^2\bigg)\;.
\eea

The near horizon geometry is a warpped product of a deformed two dimensional sphere and AdS$_3$ space. Due to  $2\pi\sqrt{\epsilon}$ periodicity in $\psi$, the near horizon geometry describes a locally AdS$_3$
with the unit radius. Besides the pinching, which does not introduce a curvature singularity, the
geometry (\ref{AdS3}) is smooth everywhere
\footnote{Although $x=0$ is a singular point but note that $x \in [\eta_3^2, \eta_4^2]$.}.

It is straightforward to check that the above geometry is a solution to the U(1)$^3$ supergravity (\ref{sugra}). In \cite{Sadeghian:2014tsa} we classified all solutions to the five dimensional gauged-supergravity with SO(2,2) isometry.  To compare metric (\ref{AdS3}) with the generic solution in \cite{Sadeghian:2014tsa}, we apply following coordinate transformations
\be
 x=\frac{\eta_3^2\eta_4^2}{\eta_3^2+(\eta_4^2-\eta_3^2)\cos^2\theta},\quad q_i=\frac{\eta_3^2\eta_4^2}{s_i(\eta_4^2-\eta_3^2)-\eta_3^2}\;.
\ee

  In \cite{Johnstone:2013eg} the EVH limit of charged rotating black hole solutions to U(1)$^3$ gauged supergravity is studied.  It has been shown that the limit contains two branches corresponding to stationary and static cases.  For the static one, where the angular momenta vanish at EVH limit, the near horizon geometry of the five dimensional solution is not well-defined. To get a regular geometry one needs to uplift the solution to higher dimensions. This is a rather generic property of all EVH static black hole solution \cite{Johnstone:2013eg}. For static black hole solutions with spherical horizon one can argue that vanishing horizon limit turn the solution to a singular geometry. However, if the solution can be uplifted to higher dimension there is a possibility to get the well-defined near horizon geometry in higher dimensions. In this case the shirking cycle comes from higher dimensions. This gives a room to find a regular geometry at the near horizon. 

The black ring solutions (\ref{solution}) have S$^2\times$ S$^1$ horizon topology and as we can see from (\ref{Shrink}), the shrinking cycle is the S$^1$ part of the horizon which lies in five dimensions and we get the well-defined near horizon geometry in five dimensions.

\subsubsection{Near-EVH near Horizon Geometry}
Near-EVH black rings  are excitations of the EVH vacua, therefore we expect them to be encoded in the near-horizon
geometry as pinching BTZ black holes. Indeed, as shall see in more detail in the next section
these excitations are described by mass and angular momentum of the pinching BTZ.
These expectations can be verified when we take $\delta \sim \epsilon$ in taking near horizon limit,
\be
\label{scale}
 y=(\rho^2-\rho_1^2-\rho_2^2)C^2\epsilon^2, \quad \rho_{1,2}=\frac{ \alpha_{1,2}}{{C}},\qquad dt=\frac{K d\tau}{ \epsilon},\qquad \phi_1=\frac{\psi}{\epsilon}\; , 
 \ee
 the metric expansion is
 \bea\label{BTZ}
&& ds^2=\frac{4(q_1q_2q_3 )^{2\over 3} {\mathcal H}(x)^{1\over 3}}{\mu^2\eta_3^2\eta_4^2 x^{4\over 3}}\bigg[-\frac{\ (\rho^2-\rho_1^2)(\rho^2-\rho_2^2)d\tau^2}{
\rho^2}+\frac{\rho^2 d\rho^2}{(\rho^2-\rho_1^2)(\rho^2-\rho_2^2)}
\nonumber \\
&& \hspace{10mm}
+ \rho^2(d\psi -
\frac{\rho_1\rho_2 d\tau}{ \rho^2}  )^2\bigg]
+\frac{ (q_1q_2q_3 )^{2\over 3}{\mathcal H}(x)^{1\over 3}}{ \mu^2 x^{10\over 3} (x-\eta_3^2)(\eta_4^2-x)} \bigg(dx^2 
\nonumber \\
&& \hspace{10mm}+{\frac{4 x^4(x-\eta_3^2)^2(\eta_4^2-x)^2 {\mathcal H}(\eta_3^2)}{\eta_3^8(\eta_4^2-\eta_3^2)^2{\mathcal H}(x)}}d\phi_2^2\bigg)\;.
\eea
The result is the same as we found in (\ref{AdS3}) but the AdS$_3$ part of the geometry is replaced by a (pinching) BTZ. Excitations appear only on the AdS$_3$ and 2d part of the near horizon geometry is intact.
\subsubsection*{Temperature $\&$ entropy: }
In order to use thermodynamics laws satisfied by BTZ
black holes, we need to compactify our five dimensional theory to three dimensions. This is achieved by considering the ansatz
\bea
ds^2=\frac{4(q_1q_2q_3 )^{2\over 3} {\mathcal H}(x)^{1\over 3}}{\mu^2\eta_3^2\eta_4^2 x^{4\over 3}} ds_3^2+ \frac{ (q_1q_2q_3 )^{2\over 3}{\mathcal H}(x)^{1\over 3}}{ \mu^2 x^{10\over 3} (x-\eta_3^2)(\eta_4^2-x)} \bigg(dx^2 
+{\frac{4 x^4(x-\eta_3^2)^2(\eta_4^2-x)^2 {\mathcal H}(\eta_3^2)}{\eta_3^8(\eta_4^2-\eta_3^2)^2{\mathcal H}(x)}}d\phi_2^2\bigg), \nonumber
\eea
and plugging it into the action of U(1)$^3$ supergravity, focusing on its Einstein-Hilbert term 
\be
\frac{1}{16\pi G_5}\int d^{5}x \sqrt{-g_5} {\mathcal R}_5 +\cdots = \frac{1}{16\pi G_3} \int d^3x \sqrt{-g_3} {\mathcal R}_3 +\cdots\;,
\ee
where $\mathcal R_5$ and $\mathcal R_3$  are Ricci scalars for the five and three dimensional metrics $ds^2$ and $ds_3^2$ respectively. We read the 3d Newton coupling constant $G_3$.
\be\label{G3}
G_3=\frac{\mu^3 \eta_3^7\eta_4^3}{8\pi q_1 q_2 q_3 \sqrt{\mathcal{H}(\eta_3^2)}}\; G_5 \;.
\ee
Thus, the temperature and entropy of the pinching BTZ black holes (\ref{BTZ}) are given by
\be\label{BTZST}
 T_{BTZ}=\frac{\rho_1^2-\rho_2^2}{2\pi \rho_1}=\frac{K}{\epsilon} T_H +\mathcal{O}(\epsilon),\qquad        S_{BTZ}=\frac{\pi\epsilon \rho_1}{2G_3}=S+\mathcal{O}(\epsilon^2)\; ,
\ee
where $T_H$ and $S$ are original black ring temperature and entropy, given by (\ref{Temp}) and (\ref{Entropy}). As expected, the BTZ entropy matches the original black ring entropy whereas its temperature agrees with the scaling of time in (\ref{scale}). 

Angular momentum and horizon angular velocity of the emergent BTZ black hole are\footnote{Note that angular direction $\psi\in [0,2\pi\epsilon]$ is $pinching$ now and therefore mass, angular momentum and the entropy of BTZ appear with a factor $\epsilon$ in front. }
\bea
\label{BTZJ}
 J_{BTZ}=\frac{\epsilon \rho_1 \rho_2}{4G_3}=\frac{J}{\epsilon}+\mathcal{O}(\epsilon),\,\,\quad \Omega_{BTZ}=\frac{\rho_2}{\rho_1}=K \Omega +\mathcal{O}(\epsilon^2),
\eea
where $J$ and $\Omega$ are defined by (\ref{J}) and (\ref{om}), evaluated at the EVH point.

This confirms the expectations when interpreting the near-EVH temperature and entropy (\ref{temp-ent}) as BTZ thermodynamic quantities.

BTZ mass is given by
\be M_{BTZ}=\frac{\epsilon}{8G_3}(\rho_1^2+\rho_2^2)\, . \ee

Comparing it with the original black ring mass reveals an interesting relation between the first law in three and five dimensional theories. We will discuss it in the next section.

\section{First law analysis, 5d vs. 3d}
In this section we analyse the first law of thermodynamics for the class of EVH-black rings we studied in the pervious sections.  We want to see how angular momentum and the dipole charges appears in the first law in the EVH limit and study whether our general first law analysis \cite{Johnstone:2013ioa} applies to the near-EVH black ring case.

To do that let us expand black ring thermodynamics quantities, such as mass, dipole charges and conjugate momenta as well as $\mathcal{P}$ and $\mathcal{A}$ appear in the first law of the black ring thermodynamics (\ref{firstlaw2}), for the small perturbation above EVH point denoted by $\epsilon$. The entropy and Hawking temperature has the following expansions
\be
S=\sum_{n=1}^{\infty} S^{(n)}\epsilon^n,\qquad T_H=\sum_{n=1}^{\infty} T^{(n)}\epsilon^n,
\ee
while the other quantities have following generic form
\be
Z=\sum_{n=0}^{\infty} Z^{(2n)}\epsilon^{2n},
\ee
where coefficients of expansions are functions of $\mu, q_i, \eta_{3,4} $ and $\alpha_{1,2}$ (or $\rho_{1,2}$ equivalently). 

Plugging these expansions in the first law (\ref{firstlaw2}), it reduces to an infinite series of equations corresponding to each oder in $\epsilon$ expnassion. Noting that in the expansion of angular momentum J we get $J^{(0)}=0$. In \cite{Johnstone:2013ioa} it is argued that for a generic EVH black hole, under certain general assumptions such as analyticity of thermodynamic quantities, we expect that the angular momentum along the pinching direction vanishes in the EVH limit.  This is indeed what we have for $\phi_1$ direction of the black rings solutions and the angular momentum along it $J$ vanishes in the EVH limit. Therefore, at the zeroth order of $\epsilon$ we get 
\be
d{M}_{e}=  \Phi_i^{(0)} dD_i^{(0)} + {\mathcal P}^{(0)} d{\mathcal A}^{(0)},\qquad \Phi_i^{(0)} =\frac{\partial M}{\partial D_i^{(0)}}|_{ext},\;\;  {\mathcal P}^{(0)}=\frac{\partial M}{\partial {\mathcal A}^{(0)}}|_{ext}\;.
\ee 
This is the BPS condition for the extremal black ring. 

The next order equations, which correspond to adding excitations to the EVH system, turns out to be
\be
d{M}^{(2)}=T_H^{(1)}dS^{(1)} +\Omega^{(0)} d J^{(2)} + D_i^{(0)} d\Phi_i^{(2)}+ D_i^{(2)} d\Phi_i^{(0)} + {\mathcal P}^{(0)} d{\mathcal A}^{(2)}  + {\mathcal P}^{(2)} d{\mathcal A}^{(0)},
\ee 
where  from (\ref{BTZST}) and (\ref{BTZJ}), we get
\be
\label{2ndterm}
\frac{1}{\epsilon K}\; T_{BTZ} dS_{BTZ}=T_H^{(1)}dS^{(1)}, \quad  \frac{1}{\epsilon K}\;\Omega_{BTZ}dJ_{BTZ}= \Omega^{(0)} dJ^{(2)}\, . 
 \ee
It is a straightforward exercise to check that the mass of near-EVH black ring and the emerged BTZ black hole satisfy the following equation 
\be
d{M}^{(2)} =  \frac{dM_{BTZ}}{\epsilon K}+\Omega^{(0)} d J^{(2)} + D_i^{(0)} d\Phi_i^{(2)}+ D_i^{(2)} d\Phi_i^{(0)} +{\mathcal P}^{(0)} d{\mathcal A}^{(2)}  + {\mathcal P}^{(2)} d{\mathcal A}^{(0)}\,.
\ee
To get this equation one should note that the variation of  three dimensional Newton constant $G_3$ should be set to zero. 
These results along with (\ref{2ndterm}) imply that whenever the EVH black ring obeys the first law of thermodynamics, emerging BTZ  also satisfies the first law of thermodynamics for BTZ black holes, i. e.  
\bea
&& d\tilde{M}=T_HdS+\Omega dJ +\sum_{i=1}^3\Phi_{D_i}dD_i + \mathcal{P} d\mathcal{A}\; \Longrightarrow \nonumber \\
&& dM_{BTZ}=T_{BTZ}dS_{BTZ} + \Omega_{BTZ}dS_{BTZ},
\eea
where we  dropped all vanishing sub-leading terms in the $\epsilon\rightarrow0$ limit.  Although we have explicitly shown  that in the near horizon region the EVH and the near-EVH black rings develop AdS$_3$ and BTZ factors, from the above analysis of the first law solely, one should expect to find an AdS$_3$ throat in the near-horizon of EVH black rings, and that this AdS$_3$ factor turns to a BTZ black hole for near-EVH cases. 
\section{The dual EVH/CFT formulation}
In this section we study a possible connection between the 2d CFT appearing in the EVH/CFT correspondence
 and the 2d chiral CFT proposed in the Kerr/CFT correspondence, studied  in section 2. Our proposal is that the 2d chiral CFT of Kerr/CFT is nothing but the Discrete
Light-Cone Quantisation (DLCQ) of a standard 2d CFT \cite{Balasubramanian:2009bg}. This has been studied for case of static and rotating charged AdS$_5$ EVH black hole solutions in \cite{Johnstone:2013eg}.

The  central charge of the 2d CFT describing the
gravitational black rings can be obtained from the standard
AdS$_3$/CFT$_2$ taking into account the pinching periodicity
\be\label{EVHC}
c=\frac{3\epsilon}{2G_3}=\frac{12\pi q_1 q_2 q_3 \sqrt{\mathcal{H}(\eta_3^2)}}{\mu^3\eta_3^7\eta_4^3 G_5}\; \epsilon\;.
\ee

To keep the central charge finite and having a finite gap in this 2d CFT, $G_5$ should scale the same as horizon area and Hawking temperature 
\be
\mathrm{horizon\;area} \sim T_H \sim G_5 \sim \epsilon \rightarrow 0\;.
\ee

It implies the black ring entropy (\ref{temp-ent}) remains finite in this limit. From (\ref{G3}) and (\ref{BTZST}) we find that the
same holds for the excitations $M_{BTZ}$ and  $J_{BTZ}$.

The quantum numbers of the associated 2d CFT $(L_0, \bar{L}_0)$ turn out to be
\be
L_0-\frac{c}{24}=\frac{\pi \eta_4 (\alpha_1+\alpha_2)^2 {\mathcal{H}^{3\over 2}(\eta_3^2)} }{2G_5 \mu^3\eta_3^{11}(\eta_4^2-\eta_3^2)^2}\;\epsilon ,\quad \bar{L}_0-\frac{c}{24} \frac{\pi \eta_4 (\alpha_1-\alpha_2)^2 {\mathcal{H}^{3\over 2}(\eta_3^2)} }{2G_5 \mu^3\eta_3^{11}(\eta_4^2-\eta_3^2)^2}\;\epsilon\;.
\ee
Using Cardy's formula we get
\be
S=2\pi \sqrt{\frac{c}{6}\left(L_0-\frac{c}{24}\right)}+2\pi \sqrt{\frac{c}{6}\left(\bar{L}_0-\frac{c}{24}\right)}=\frac{4\pi^2\sqrt{q_1q_2q_3 }\;{\mathcal{H}}(\eta_3^2)\;\alpha_1 }{\mu^3\eta_3^9\eta_4(\eta_4^2-\eta_3^2)G_5}\;\epsilon ,
\ee
we observe that the near-EVH black ring entropy (\ref{temp-ent}) is reproduced.  

To compare this result with Kerr/CFT we note that the Kerr/CFT applies for extremal finite size black rings,
while EVH/CFT works for near-EVH black ring which can be non-extremal. Therefore
we must compare them in a region of parameters  where both apply. This can be
done by considering the extremal excitations in the EVH/CFT side 
and restricting to the vanishing entropy limit in the Kerr/CFT side. The latter step is subtle and may involve
singular limits. 

On the bulk side we should note that taking the near horizon limit of near-EVH black rings does not commute with taking the near-EVH limit of the near horizon geometry of  extremal finite horizon black rings. These two limits lead to 
different geometries. On the CFT side, to reproduce the appearance of a vanishing cycle to account for
the vanishing entropy the Kerr/CFT central charge tends to zero. 

Despite these facts, we observe that the Kerr/CFT central charge (\ref{KC}) associated with the vanishing U(1) isometry cycle $\tilde{\phi}_1$ remains finite in the EVH limit and the result matches the AdS$_3$ Brown-Henneaux central charge (\ref{EVHC}) computed in the EVH/CFT correspondence.

To do this end, we note that at the EVH limit the leading term in the Kerr/CFT central charge is given by
\be
c_{\tilde{\phi}_1} =\frac{12\pi q_1 q_2 q_3 \sqrt{\mathcal{H}(\eta_3^2)}}{\mu^3\eta_3^7\eta_4^3 G_5} +\mathcal{O}(\epsilon^2)\; ,
\ee
From coordinate scaling (\ref{scale}), the central charge scales as
\be
c_{\psi}=\epsilon\; c_{\tilde{\phi}_1},
\ee
which matches with central charge (\ref{EVHC}).  Within the Kerr/CFT proposal  one
may then expect that in the near horizon of  the extremal rings we have a chiral CFT
description associated with the EVH/CFT via the DLCQ description.  
\section{Conclusion}
In this work we analysed the near horizon geometry of a recently found dipole electric charged black ring \cite{Lu:2014afa}. We have shown that the black ring parameter space contains a region in which an AdS$_3$ factor emerges in the near horizon geometry. This happens when the Hawking temperature and black ring's horizon area are small and the same order. We have shown that there are two possibilities to get this limit corresponding to static or stationary limits. In the stationary limit all dipole charges vanish and the near horizon limit is similar to the neutral single rotating black ring.  In the static limit where angular momentum vanishes, the AdS$_3$ part of the near horizon turns to  BTZ when we turn on infinitesimal non-extremalities while keeping the ratio of temperature to area of horizon fixed.  We used AdS$_3$/CFT$_2$ to find statistical entropy of black ring solution and studied its relation to Kerr/CFT. The black ring solution at the generic point in parameter space has conical singularity. We have studied the first law of thermodynamics for this black ring solution and showed how to modify it to account for the effect of conical singularity. We have also studied the first law when temperature and horizon area are both small. By expanding both sides we have shown that it reduces to the first law for BTZ, therefore the appearance of BTZ in the near horizon geometry is predicted indirectly in the first law analysis. 
 \section*{Acknowledgement}
 We would like to thank M. M. Sheikh-Jabbari and J. Vazquez-Poritz for their collaboration at the early stage of this work. We would like to thank E. \'O Colg\'ain and H. Lu for discussion and their insightful comments.
\appendix
\section{Five-dimensional Myers-Perry black hole} 
The simplest example of a black hole with one angular momentum in five dimensions is Myers-Perry black hole solution. The metric in Boyer-Lindquist type coordinate is given by
\bea 
\label{MPBH}
ds^2 & = & -dt^2 +\Sigma\,\left(\frac{r^2}{\Delta}\,dr^2 + d\theta^2 \right)
+r^2\,\sin^2\theta \,d\phi^2
+(r^2 + a^2)\,\cos^2\theta \,d\psi^2 \nonumber \\
&  & +\,\frac{m}{\Sigma}\, \left(dt 
- a\,\cos^2\theta \,d\psi \right)^2,
\eea

where
\begin{eqnarray}
\Sigma=r^2+ a^2 \,\sin^2 \theta, \;\;\;\;\;\;
\Delta= r^2 (r^2 + a^2 -m),
\end{eqnarray}
and $\,m \,$  is a parameter related to the physical
mass of the black hole, while the parameters $\,a \,$
is associated with its angular momentum. This metric 
depends only on two coordinates, $0< r < \infty$ and $0\le\theta\le\pi/2$, 
and it is independent of time, $-\infty < t < \infty$, and the azimuthal 
angles, $0 < \phi,\psi < 2\pi$. The inner and outer horizons are located at non-negative zeroth of function $\Delta$
\begin{eqnarray}
r_-=0, \quad r_+=\sqrt{m-a^2}\;.
\label{hradius}
\end{eqnarray}
Notice that the horizon exists if and only if
$
|a| \,\leq m
$
otherwise, the metric describes a naked singularity.
Hawking temperature is given by 
\be 
T_H=\frac{\sqrt{m-a^2}}{2\pi m},
\ee 
 ADM mass, angular momentum and horizon angular velocity are given by
\begin{eqnarray}
\label{charges-Kerr-5D}
M_0=\frac{3\, \pi\, m}{8\, G_5}, \;\; J_{\psi}=\frac{\pi\, m\, a}{4\, G_5},\quad \Omega_{\psi}=\frac{a}{m}\;.
\end{eqnarray}
Area of the horizon is given by following integral
\be 
A=\int_0^{2\pi}d\phi \int_0^{2\pi}d\psi \int_0^{2\pi}d\theta \sqrt{g_{\theta\theta}(g_{\phi\phi}g_{\psi\psi}-g_{\phi\psi}^2)}=2\pi^2 m \sqrt{m-a^2}\;.
\ee 
Therefore the Bekenstein-Hawking entropy is 
\be 
S_{BH}=\frac{A}{4G_5} = \frac{\pi^2 m \sqrt{m-a^2}}{2G_5}\;.
\ee  
\subsection*{EVH point and Static BTZ}
The EVH limit is defined by $T\rightarrow 0$ and $S_{BH}\rightarrow 0$ at the same time. In addition one may also wish to keep the ratio $\frac{S_{BH}}{T}$ finite. For the Myers-Perry black hole solution this is given by $r_+\rightarrow 0$ while keeping $m$ finite. We introduce 
\be
 a\simeq \sqrt{m}-\alpha\epsilon^2 \;.
\ee
The near horizon limit is obtained by taking the limit $\epsilon \rightarrow 0$ and performing following change of coordinate
\be
r= \epsilon \sqrt{m} \rho, \;\;\;\;\;\;\;\; \phi =\frac{\tilde{\phi}}{\epsilon} , \;\;\;\;\;\;\;\; \psi=\tilde{\psi} - \frac{t}{\sqrt{m}}, \;\;\;\;\;\;\;\;  t=\frac{\sqrt{m}\tau}{\epsilon}\;.
\ee
The result is
\bea
ds^2=m\sin^2\theta \bigg[-(\rho^2-\rho_+^2) d\tau^2+\frac{d\rho^2}{\rho^2-\rho_+^2} + \rho^2 d\tilde{\psi}^2 +\frac{\sin^2\theta}{\cos^4\theta} d\tilde{\phi}^2 +d\theta^2 \bigg],
\eea
where 
\be 
\rho_{+}^2=\frac{2\alpha}{\sqrt{m}}\;.
\ee 
To get real $\rho_{\pm}$ we must impose $\alpha>0$. For the emerged static BTZ, we get
\bea
&& M_{BTZ}=\frac{\epsilon \alpha}{4G_3 \sqrt{m}}\quad T_{BTZ}=\frac{\sqrt{2\alpha}}{2\pi m^{1/4}},\quad S_{BTZ}=\frac{\pi\epsilon \sqrt{2\alpha}}{2G_3 m^{1/4}},
\eea

{}

\end{document}